\documentclass[aps,preprint,epsfig]{revtex4}
\usepackage{graphicx}
\begin{document}
\title{Approximate solutions for the Skyrmion}
\author{J. A. Ponciano, L. N. Epele, H. Fanchiotti and C. A. Garc\'{\i}a Canal}
\address{Departamento de F\'{\i}sica, Universidad Nacional de La Plata\\ C.C. 67 (1900) La Plata \\ Argentina \\ }

\address{\rm{In memoriam Goyi Mignaco\\}}

\begin{abstract}
We reconsider the Euler-Lagrange equation for the Skyrme model in the hedgehog ansatz and study the analytical properties of the solitonic solution. In view of the lack of a closed form solution to the problem, we work on approximate analytical solutions.

We show that Pad\'e approximants are well suited to continue analytically the asymptotic representation obtained in terms of a power series expansion near the origin, obtaining explicit approximate solutions for the Skyrme equations. We improve the approximations by applying the 2-point Pad\'e approximant procedure whereby the exact behaviour at spatial infinity is incorporated. An even better convergence to the exact solution is obtained by introducing a modified form for the approximants. The new representations 
share the same analytical properties with the exact solution at both small and large values of the radial variable $r$.

\hfill

\noindent{\it PACS number(s): 11.10.Lm, 12.39.Dc, 02.30.Mv}
\end{abstract}
\maketitle

\section{Introduction}

The Skyrme model \cite{Skyrme} provides a picture of baryons interacting via meson exchanges \cite{Witten}, \cite{'t Hooft}.
The model is based on the pre-QCD nonlinear $\sigma$ model, with 
the same group of chiral symmetry $G=SU(N_f)\times SU(N_f)$ \cite{Adkins}. Classical
stability arguments require the presence of an additional term in the
nonlinear $\sigma$-model Lagrangian \cite{Derrick}; this term was introduced initially by Skyrme.
In this scenario, baryons emerge as solitonic solutions of the
Euler-Lagrange equations \cite{Rajaraman}, which are  highly nonlinear second order equations and cannot be solved in exact closed form. Solutions are obtained by numerical integration. This is an awkward situation for the evaluation of physical quantitie
s.  In practice it is more convenient to have explicit approximate representations for the solitonic solutions.

In the present work we obtain reliable approximate representations for
the solution of the Skyrme model in the $SU(2)_L \times SU(2)_R$ sector by
means of the Pad\'e Approximant (PA) \cite{Baker} procedure whereby the series
expansion solution of the differential equations is continued analytically.

The successful implementation of PA summation relies on the simple
structure of the series solution near the origin which is almost an
alternate geometric series.
In order to reproduce the essential features of the exact solution at
large distances, we use the 2-point PA method and construct modified
representations that share the exact properties of the skyrmion
configuration near both boundaries. Futhermore our approximate solutions
are reliable in the whole interval of the independent variable $r$, namely $r=0$ to infinity.

This paper is organized as follows. In section II we introduce the Skyrme model in the $SU(2)_L\times SU(2)_R$ sector and restrict the problem using the hedgehog ansatz. In section III we present properties of the hedgehog solution showing  power series r
epresentations for the {\it Chiral Angle}. In section IV we introduce and use our approach based on the PA method. Finally, in section V we draw the conclusions of our work.

\section{The Skyrme model}

Following Skyrme's proposal, baryons are soliton in the $\sigma 
-$nonlinear model with an additional stabilizing term. We will restrict
ourselves to the sector $SU(2)_L\times SU(2)_R$ where the Lagrangian
density of the model reads
\begin{equation}\label{modelskyrme}
\it{L}=\frac{\it{f}_\pi ^2}{4}Tr(\partial _\mu U \partial ^{\mu}
U)+\frac{1}{32g^2} Tr[U^{+} \partial _{\mu} U, U^{+} \partial _{\nu} U]^2,
\end{equation}
being $U$ a unitary operator. The parameter $f_{\pi}$ is the usual pion decay constant whose experimental value is $93$ MeV and $g$ is the dimensionless Skyrme parameter.
Equivalently, $U$ can be expressed in terms of a scalar and a pseudoscalar field, $\sigma$ and $\mbox{\boldmath $\pi$}$ respectively, in the following form
\begin{equation}
U=(\sigma + i\mbox{\boldmath $\tau \cdot \pi$})/f_{\pi}
\end{equation}
It should be stressed that $\sigma$ and $\mbox{\boldmath $\pi$}$ are not independent fields, and are related through $\sigma ^2 +\mbox{\boldmath $\pi$}^2=f_{\pi}^2$. This restriction is the source of nonlinearities in the theory.

Rewriting $\mbox{\boldmath $\pi$}$ and $\sigma$ in terms of a new field $\mbox{\boldmath $F$}=f_{\pi}F \mbox{\boldmath $\hat{\varphi}$}$
\begin{equation}\label{piones}
\sigma=f_{\pi}\cos F, \qquad
\mbox{\boldmath $\pi$}=f_{\pi}\mbox{\boldmath $\hat{\varphi}$} \sin F,
\end{equation}
and going back to $U$, we obtain the representation
\begin{equation}
U=\exp(i\mbox {\boldmath $\tau \cdot F $})
\end{equation}
Previous analysis \cite{Mattis} have shown that a reasonable agreement with experimental values of physical quantities is obtained when the hedgehog configuration of $U$ is adopted, this is a particlular static ansatz where $\mbox{\boldmath $F$}$ takes a 
spherically symmetric form, namely
\begin{equation}
U_s(\tilde r)=\exp(i\mbox{\boldmath $\tau \cdot \hat n$ } F(\tilde r))
\end{equation}
where $\mbox{\boldmath $\hat n$}$ denotes a unit radial vector.

In this representation, the energy reduces to a functional of the  
{\it Chiral Angle}, $F(\tilde r)$, alone
\begin{equation}\label{energia}
E[F]=2\pi \int _0 ^\infty d\tilde r \left[
f_{\pi}^2\tilde r^2\left(\frac{dF}{d\tilde r}\right)^2 +2f_{\pi}^2\sin^2F
+\frac{1}{g^2}\sin^2F\left(2\left(\frac{dF}{d\tilde r}\right)^2+\frac{\sin^2F}{\tilde r^2}
\right)\right].
\end{equation} 

The variation $\delta E /\delta F =0$ generates a nonlinear differential
equation for $F(\tilde r)$, whose solution corresponds to the static
configuration of minimal energy. Written in terms of the dimensionless
variable $r=gf_{\pi}\tilde r$, the differential equation for $F(r)$
reads
\begin{eqnarray}\label{ecuacion}
[r^2 + 8\sin^2F(r)]\frac{d^2F(r)}{dr^2}&=&\frac{4}{r^2}\sin^2F(r)\sin[2F(r)]
+
\sin[2F(r)]-2r\frac{dF(r)}{dr} \nonumber \\
& &-4\sin[2F(r)]\left[\frac{dF(r)}{dr}\right]^2.
\end{eqnarray}
As finite energy solutions are required, $U(r)$ must tend to an
arbitrary constant element of $SU(2)$ at spatial infinity. Choosing 
$U(r)\rightarrow 1$ as $r\rightarrow \infty$ , implies the boundary 
condition
\begin{equation}
F(\infty)=0
\end{equation}

This condition defines a mapping $S^3 \rightarrow S^3$, from the compactified 
configuration space to the identity in the target space in $SU(2)$ which is isomorfic to $S^3$, falling into
distinct equivalence classes labeled each by the winding number Z
\begin{equation}
\Pi_3(SU(2)) =\Pi_3(S^3)=Z, 
\end{equation}
The integer number Z, that counts the coverings of the target space, is 
defined as the topological charge $q=\int dx^3 B_0 $ related to the
baryonic current 
\begin{equation}
B_{\mu} = \frac{1}{24\pi^2}\epsilon_{\mu \nu \rho \sigma}
Tr[(U^{+}\partial ^{\nu} U) (U^{+}\partial ^{\rho} U)  (U^{+}\partial
^{\sigma}U)]
\end{equation}
For the hedgehog ansatz, the baryon number charge density is
\begin{equation}
B_{0}=-\frac{1}{2\pi ^2} \frac{dF}{dr} \frac{\sin^2 F}{r^2}
\end{equation}
and corresponds to a baryon number
\begin{equation}
B=\frac{1}{2\pi} \left[2F(0)-2F(\infty)-\sin 2F(0)+\sin 2F(\infty)\right]
\end{equation}
From the foregoing, it follows that the solution belonging to
the sector with baryon number $B=1$, satisfies the condition    
\begin{equation}
F(0)=\pi .
\end{equation}
The numerical solution labeled with $B=1$ is shown in Fig.1 .  Inserting
this solution into Eq. (\ref{energia}) yields $E=23.2 \pi f_{\pi}/g$.  

\section{Properties of the hedgehog solution}

Before applying the PA procedure to the Skyrme model, we
shall consider the solution of the Euler-Lagrange equation at both ends of the definition domain.
Near the origin, a power series solution can be obtained \cite{Mignaco},
\begin{equation}\label{origin}
S_{H}(r)=\pi + F_1 \tilde r+\frac{1}{3!} F_3 \tilde r^3 + \frac{1}{5!} F_5 \tilde r^5 + ...\ ,
\end{equation}
where $F_1$ turns out to be undetermined. All even powers have vanishing
coefficients. Odd power coefficients are written in terms of $F_1$ and
the dimensionless parameter $\phi$, defined as $\phi = F_1 /gf_{\pi}$; for
example, 
\begin{equation}\label{F3}
F_3=-\frac{4}{5}F_{1}^{3} \frac{1+2\phi ^2}{1+8\phi ^2}
\end{equation}
\begin{equation}\label{F5}
F_5=\frac{24}{7}F_{1}^{5} \frac{1+(32/5)\phi ^2 + (88/5) \phi ^4 + (488/5)
\phi ^6}{1+24 \phi ^2 +192 \phi ^4 + 512\phi ^6},
\end{equation}
while the results for higher terms may be found elsewhere \cite{Mignaco}. 
In terms of the dimensionless variable $r=gf_{\pi}\tilde r$, expansion 
(\ref{origin}) becomes
\begin{equation}\label{adimensional}
S_{H}(r)=\pi +\phi r+\frac{1}{3!} \phi_3 r^3 + \frac{1}{5!} \phi_5 r^5 +
...,
\end{equation}
where $\phi_3$ and $\phi_5$ are given by the same expressions as $F_3$ and
$F_5$
but with $\phi$ instead of $F_1$.  From now on, we shall be concerned only 
with the dimensionless variables.

For large values of $r$ , there is a solution of the form $C/r^2$,
as it can be readily seen from trying this form in equation
(\ref{ecuacion}) and keeping only linear terms  in $F(r)$.

In order to find contributions of higher order in $1/r$, it is appropriate
to perform the change of variable $\rho = 1/r$ in equation (\ref{ecuacion}). 
In this new variable it reads:
\begin{eqnarray}\label{infinitequation}
[\rho ^2 +8\rho ^4 \sin F(\rho)]\frac{d^2 F(\rho)}{d\rho ^2}&+&8 \rho
^{3} \sin^{2}F(\rho)\frac{F(\rho)}{d\rho}
 +4\rho ^{4}\sin^{2}[2F(\rho)] \left[
\frac{dF(\rho)}{d\rho}\right]^2 \nonumber \\
&& -4\rho ^2 \sin^{2} F(\rho)\sin[2F(\rho)]-\sin [2F\rho]=0
\end{eqnarray} 
Trying a power series expansion as before,
\begin{equation}\label{infinity}
F(\rho)=K_0 +K_1 \rho +\frac{1}{2!} K_2 \rho ^2 +\frac{1}{3!} K_3 \rho
^3+...  
\end{equation}
we find that the series solution contains just even powers. The first few
coefficients written in terms of $K_2$ are \cite{Mignaco}
\begin{eqnarray}
K_4&=&0 \\
K_6&=&-\frac{30}{7} K_2^{3}\\
K_8&=&-6720K_2^{3}
\end{eqnarray}
Just as the initial slope $\phi$, the parameter $K_2$ cannot be determined
from the differential equation. Approximate values for $\phi$ and $K_2$
are given in Ref.\cite{Mignaco}. These values were 
determined through the use of integration routines based on shooting 
methods \cite{Adkins}. This is some sort of eigenvalue problem for $\phi$ and $K_2$. The solution is obtained by numerical integration of the differential equation, starting from the origin where the value of $\phi$ must be such as to satisfy the correct 
boundary condition corresponding to the relevant solitonic solution. Once $\phi$ is found, the value of $K_2$ is obtained in similar way using equation (\ref{infinitequation}). The reported values are \cite{Ebrahim}.
\begin{equation}
\phi=-1.0037 \qquad \qquad K_2=17.2772
\end{equation}
These values, as well as the vality of the numerical solution can be checked in a consistent manner. We will not go on through this and just refer the reader to the Ref. \cite{Ananias}

Let us make a slight modification in our analysis by rewritting the power expansion (\ref{adimensional}) in the following form
\begin{equation}\label{newseries}
S_{H}(r)=\pi + \phi r + a_1 r^3\left(1+ \frac{a_2}{a_1}r^2+ \frac{a_2}{a_1}\frac{a_3}{a_2}r^4+\ldots \right),
\end{equation}
where the coefficients $a_1$, $a_2$, $a_3$, \ldots are defined through
\begin{equation}
a_1=\frac{\phi_3}{3!}, \qquad a_2=\frac{\phi_5}{5!} \qquad a_3=\frac{\phi_7}{7!},
\end{equation}
and so on. As shown in table I, the numerical values of the ratios $-a_{n+1}/a_n$ vary very slowly with $n$. Thus, one can introduce the nearly constant $R\simeq -a_{n+1}/a_n$ (for any value of $n$). Consequently,  the portion within the brackets in (\ref
{newseries}) can be replaced, to a good degree of approximation, by the
alternate geometric series
\begin{equation}
1-Rr^2+(Rr^2)^2-(Rr^2)^3+(Rr^2)^4-...,
\end{equation}
which has a radius of convergence $r_0$ defined by the relation
$r^2_0=|1/R|$.

Within the radius of convergence, $r\leq r_0$, expansion (\ref{newseries}) may be replaced by
\begin{equation}
S_H\simeq \pi+\phi r+a_1r^3(1-Rr^2+(Rr^2)^2-(Rr^2)^3+(Rr^2)^4-...).  
\end{equation}
An alternate series like $1-x+x^2-x^3+...$ can be continued beyond its radius of convergence by the function $(1+x)^{-1}$, so that $S_H$ can also be analytically continued for all positives values in the real $r$ axis using the representation
\begin{equation}\label{pa11}
F(r)=\pi+\phi r+\frac{a_1r^3}{1+Rr^2}
\end{equation}
Taking $R=-a_2/a_1$, Eq. (\ref{pa11}) is exactly the $[3,2]$ order Pad\'e Approximant representation of the series (\ref{newseries}). The obvious task is to test whether higher order of approximation can provide a better representation of the soliton. 
There is a great amount of examples \cite{pade} where Pad\'e Approximant procedure provides reliable analytic continuation of power series expansions and that PA representations converge rapidly to the relevant function of the problem. 
In the next section we will build different representations of PA to the hedgehog solution showing that the features of reliability and fast convergence also hold in this problem.

\section{Pad\'e approximants to the hedgehog solution}

Our approach is based on Pad\'e approximants, which are rational functions
used to provide an
analytic continuation of a power series representation of a given
function, and are typically known to accelerate the convergence of the
series. The Pad\'e approximant $P_{[M,N]}(x)$ of order $[M,N]$ to the
series
$S(x)=\sum_n a_n x^n$ is defined as the ratio of two polynomials
\begin{equation}\label{pade}
P_{[M,N]}(x)=\frac{\sum\limits_{k=0}^{M} A_{k}x^{k}}{\sum\limits_{k=0}^{N}
B_{k}x^{k}}
\end{equation}
where we set $B_0=1$ without loss of generality. The remaining $M+N+1$
coefficients are chosen so that the first $M+N+1$ coefficients in
the Taylor expansion of $P_{[M,N]}$ coincide with the series $S(x)$
through order $M+N$. Conversely, if only the first $M+N+1$ coefficients 
of the series $S(x)$ are known, such a Pad\'e approximant can be used 
to predict the next coefficient in the series \cite{Ellis}.
 
The PA of order $[M,N]$ to the hedgehog solution is found by comparing its Taylor expansion at the origin with series (\ref{origin}). From all the possible combinations $(M,N)$ allowed by the input series expansion (\ref{origin}), we are interested in tho
se that may provide a suitable representation to the hedgehog solution. In order to enforce the conditions that fix the soliton solution to be of winding number one, we will require that the behaviour at infinity of the PA representations is given by the 
leading term of expansion (\ref{infinity}).
To this end we impose the contraint $N-M=2$.  We have computed the
values of coefficients $a_k$ and $b_k$ corresponding to the sequence fo PA's $[0,2]$, $[1,3]$ and $[2,4]$ using 
the value $\phi_1=-1.003$ found by numerical methods. The
results are displayed in figure 1. The explicit dependence of the PAs in terms of the coefficients of series (\ref{adimensional}) is
\begin{eqnarray}
F_{[0,2]}&=&\frac{\pi}{1-\frac{\phi r}{\pi}+\left(\frac{\phi r}{\pi}\right)^2}, \nonumber \\
F_{[1,3]}&=&\frac{\pi+\left(\phi+\frac{\phi \phi_3 \pi^2}{6 \phi^3 +\phi_3 \pi^2}\right)r}{1+\frac{\phi \phi_3 \pi}{6 \phi^3 +\phi_3 \pi^2}r-\frac{\phi^2\phi_3 }{6\phi^3+\phi_3 \pi^2}r^2-\frac{\phi_3^2 \pi}{6(6\phi^3+\phi_3\pi^2)}r^3},
\end{eqnarray}
for $[0,2]$ and $[1,3]$ respectively.

The PA representations built from the exact series solution near the origin do not guarantee the exact behaviour at intermediate and large values of $r$. However, one could use the asymptotic solution at infinity given in (\ref{infinity}) in order to find better representations of the hedgehog \cite{Linde}.
Yet, the right behaviour at infinity may be incorporated using the 2-point Pad\'e Approximant procedure. This is a natural extension of the usual PA method which consists in fixing almost one coefficient of the approximant using the asymptotic series solution at infinity \cite{Baker}. We have computed the sequence of 2-point PA of order $[1,3]$, $[2,4]$,
 $[3,5]$ (see figure 2). Certainly, the 2-point PA representations are more reliable than the ordinary ones. The explicit form of PA $[1,3]$ is 
\begin{equation}
F_{[1,3]}=\frac{\pi+\frac{K_2(6\phi^3+\phi_3\pi^2)}{6(K_2\phi^2-2\pi^3)}}{1-\frac{\pi(K_2\phi_3+12\phi\pi)}{6(2\pi^3-K_2\phi^2)}-\frac{\phi(K_2\phi_3+12\phi\pi)}{6(K_2\phi^2-2\pi^3)}+\frac{6\phi^3+\phi_3\pi^2}{3(K_2\phi^2-2\pi^3)}}
\end{equation}

The next step is to include the subleading behaviour of the exact solution at infinity within the PA representations. We have done this up to order $O(1/r^3)$ as shown in figure 3.
As we go on with the procedure, the degree of approximation is improved both near the origin and for large values of $r$. However, for intermediate values of $r$, the PA representations are slowly convergent.
Nevertheless we will show that there is a suitable functional form that provides very good representations to the hedgehog solution. 

To this end, let us prescribe a modified PA of order $[0,4]$ through
\begin{equation}\label{04m}
\tilde F_{[0,4]}(r)=\frac{a}{(1+b_{1}r+b_{2}r^2+b_{3}r^3+b_{4}r^4)^{1/2}},
\end{equation}
such that the coefficients $a_0$, $b_1$, $b_2$ and $b_3$ are determined by
expanding $\tilde F_{[0,4]}(r)$ in a power series near the origin and matching
the terms,
order by order, with those of the exact solution (\ref{adimensional}) up to
$O(r^3)$, and the remaining coefficient $b_4$ is fixed by using the leading term at infinity given by (\ref{infinity}). We found explicitly,
\begin{equation}
\tilde F_{[0,4]}(r)=\frac{\pi}{(1-\frac{2\phi}{\pi}r+\frac{3\phi^2}{\pi^2}r^2+\frac{-12\phi^3-\phi_3\pi^2}{3\pi^3}r^3+\frac{4\pi^2}{K_2^2}r^4)^{1/2}}.
\end{equation}
The form (\ref{04m}) can be generalized to higher orders $[0,k]$ as
\begin{equation}\label{0km}
\tilde F_{[0,k]}(r)=\frac{a}{(1+b_{1}r+b_{2}r^2+...+b_{k}r^k)^{2/k}}
\end{equation}
where the unknown coefficients may be found following the prescription
described for the particular case $k=4$.
We have built the representations corresponding to $k=6$, $k=8$, and $k=10$ (see Fig. 4). By construction, such representations have the exact Taylor expansion near the origin up to order $O(r^5)$, $O(r^6)$ and $O(r^7)$ respectively, as well as the exact 
leading behaviour at infinity.  The new PA representations improve in a remarkable way the representations obtained so far.

Following the program, an even better agreement for the large $r$ behaviour may be obtained by including in the modified representations (\ref{0km}) the subleading behaviour of the hedgehog solution. In figure 5 we have depicted the approximants so constr
ucted. The figure shows an exceptional good agreement betwen the numerical solution and the modified PA given by (\ref{0km}). In order to show the utility of PA procedure as a method of analytic continuation, we have included in the same figure the power 
series representation near the origin up to and including order $O(r^9)$.

The utility of PA method in this problem is outstanding, particularly for its simple and reliable application to the highly nonlinear Euler-Lagrange equation of Skyrme. Moreover, the PA representations reproduce exactly the main properties of the hedgehog
 solution at both  small and large values of $r$.

In order to further check the reliability of the PA representations, we have calculated its corresponding baryon number given by 
\begin{equation}
B=-\frac{1}{2\pi ^2} \int_0^{\infty}\frac{dF}{dr} \frac{\sin^2 F}{r^2}4\pi r^2dr,
\end{equation} 
and the minimal energy (\ref{energia}) whose numerical value is $E=72.88 f_{\pi}/g$. 
We obtain $B=1$ for the 2-point PAs which is exactly the topological number of 
the skyrmion. As for the energy, the results for the last sequence of approximants in units of $f_{\pi}/g$ are $E_{[0,4]}=73.60$, $E_{[0,6]}=73.02$, $E_{[0,8]}=72.96$, and $E_{[0,10]}=72.94$, which converge to the numerical value reported in previous work
s \cite{Adkins}.

\section{Final remarks}
We have shown that the suggested PA representations for the Skyrme solutions can incorporate in a simple way the properties of the exact solution near both physical boundaries of the problem. This solves the difficulty acquainted in Ref.\cite{Ananias}, wh
ere the authors assert the impossibility to obtain a PA to the series solution about the origin which, in addition, reproduces the exact behaviour of the chiral angle at infinity. Moreover, our representations avoid the use of two separate analytical repr
esentations for the skyrmion, one for each of the two regions, which was the common approach to this problem \cite{Iwasaki}, clearly unsuitable for phenomenological calculations.

Certainly, the results presented here highlight the utility of rational functions, PA in particular, for analytic continuation of asymptotic series. 
In this problem, the knowledge of the asymptotic behaviour at two points favors a better convergence to the relevant function. 
The approximate analytical solutions can be used with reliability in the evaluation of physical quantities. The calculations are further easier and more manageable than numerical computations.

\newpage 
\begin{table}
\begin{center}
\caption{Ratios of the coefficients of the series expansion (\ref{newseries})}
\begin{tabular}{ccccccc}
\hline
\hline
$a_2/a_1$ & $a_3/a_2$ & $a_4/a_3$ & $a_5/a_4$ & $a_6/a_5$ & $a_7/a_6$ &
$a_8/a_7$ \\
\hline
-0.102 \quad & -0.136 \quad & -0.133 \quad & -0.139 \quad & -0.147 \quad & -0.149 \quad & -0.152  \\
\hline
\hline
\end{tabular}
\end{center}
\end{table}

\begin{figure}[!h]\label{fig.1}
\centering
\includegraphics[width=9cm]{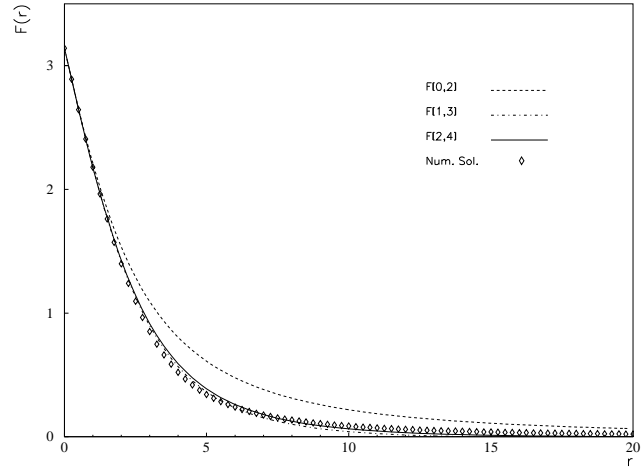}
\caption{Pad\'e Approximants to the hedgehog solution.}
\end{figure}
\begin{figure}[!h]\label{fig.2}
\centering
\includegraphics[width=9cm]{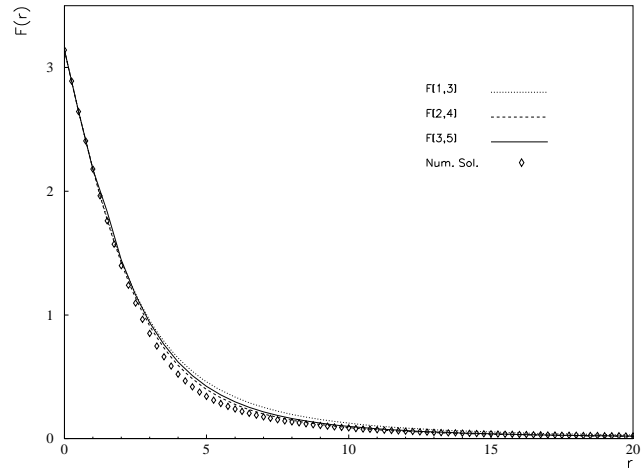}
\caption{Two point PA to the hedgehog solution.}
\end{figure}

\begin{figure}[!h]\label{fig.3}
\centering
\includegraphics[width=9cm]{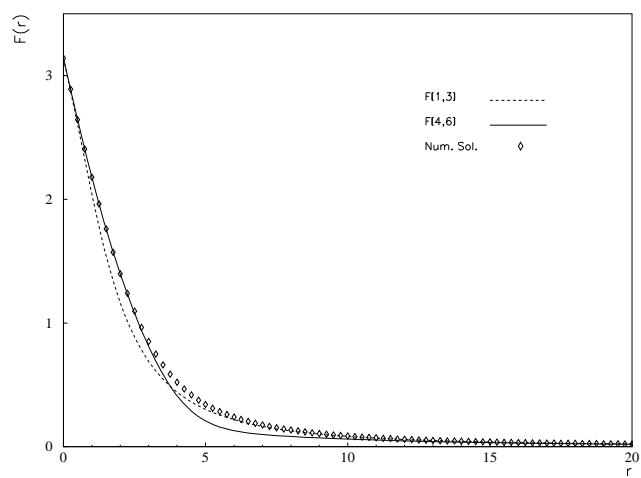}
\caption{Two point PA exact up to order $O(1/r^3)$ in the limit $r\rightarrow \infty$.}
\end{figure}

\begin{figure}[h!]\label{fig.4}
\centering
\includegraphics[width=9cm]{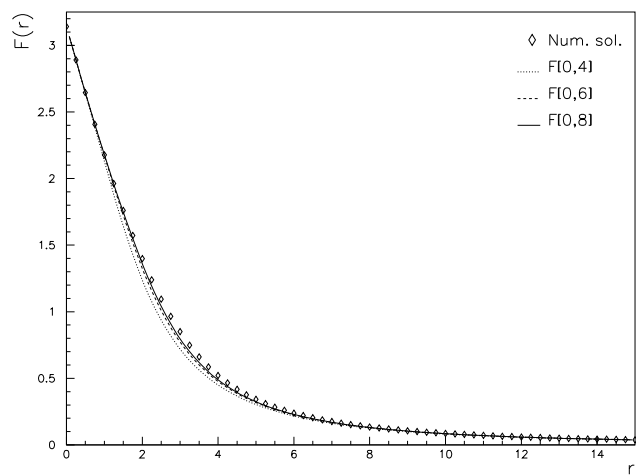}
\caption{Sequence of modified PA to the hedgehog solution.}
\end{figure}

\begin{figure}[h!]\label{fig.5}
\centering
\includegraphics[width=9cm]{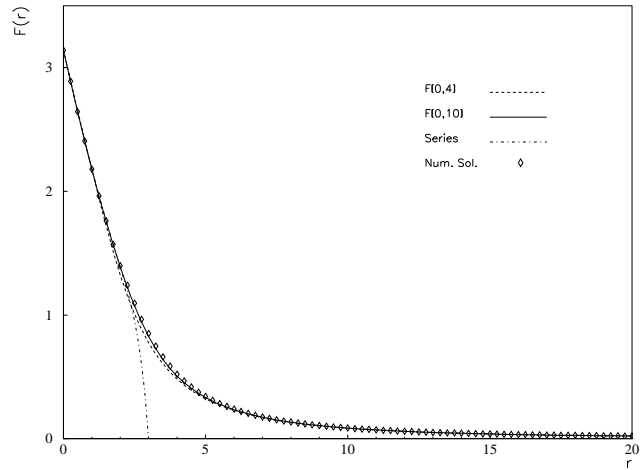}
\caption{Modified PA representations which reproduce the exact leading and subleading behaviour at $r\rightarrow \infty$.}

\end{figure}
\end{document}